\documentclass[aps,twocolumn,preprintnumbers,floats]{revtex4}

\usepackage{amssymb}
\usepackage{braket}
\usepackage{booktabs}
\usepackage{color}
\usepackage{epsfig}
\usepackage{float}
\usepackage{graphicx} % Include figure files
\usepackage[]{hyperref}
\usepackage{indentfirst}
\usepackage{longtable}
\usepackage{lscape}
\usepackage{mathrsfs}
\usepackage{multirow}
\usepackage{setspace}
\usepackage{txfonts}

\begin{document}

\title{Three body open flavor decays of higher charmonium and bottomonium}

\author{
Xin-Zhen Weng$^{1}$~\footnote{xzhweng@pku.edu.cn},
Li-Ye Xiao$^{1,2}$~\footnote{lyxiao@pku.edu.cn},
Wei-Zhen Deng$^{1}$~\footnote{dwz@pku.edu.cn},
Xiao-Lin Chen$^{1}$\footnote{chenxl@pku.edu.cn},
and
Shi-Lin Zhu$^{1,2,3}$~\footnote{zhusl@pku.edu.cn}
}

\affiliation{$^1$School of Physics and State Key Laboratory of Nuclear Physics and Technology, Peking University, Beijing 100871, China}
\affiliation{$^2$Center of High Energy Physics, Peking University, Beijing 100871, China}
\affiliation{$^3$Collaborative Innovation Center of Quantum Matter, Beijing 100871, China}

\begin{abstract}
In the present work, we study the Okubo-Zweig-Iizuka-allowed three body open flavor
decay properties of higher vector charmonium and bottomonium states
with an extended quark pair creation model. For the bottomonium
system, we get that (i) the $BB\pi$ and $B^*B^*\pi$ partial decay
widths of the $\Upsilon(10860)$ state are consistent with the
experiment, and the $BB^*\pi$ partial decay width of the
$\Upsilon(10860)$ state is smaller but very close to the Belle's
experiment. Meanwhile, (ii) the $BB^*\pi$ and $B^*B^*\pi$ decay
widths of $\Upsilon(11020)$ can reach $2\sim3$ MeV. In addition,
(iii) for most of the higher vector charmonium states, the partial
decay widths of the $DD^*\pi$ and $D^*D^*\pi$ modes can reach up to
several MeV, which may be observed in future experiments.
\end{abstract}

\maketitle

%%%%%%%%%%%%%%%%%%%%%%%%%%%%%
\section{Introduction}
%%%%%%%%%%%%%%%%%%%%%%%%%%%%%

In 2003, the Belle Collaboration reported the first observation of
the charmoniumlike state $X(3872)$ in exclusive
$B^{\pm}{\to}K^{\pm}\pi^{+}\pi^{-}J/\psi$ decays~\cite{Choi:2003ue}.
This state was later confirmed by the CDF~\cite{Acosta:2003zx},
D0~\cite{Abazov:2004kp}, {\it BABAR}~\cite{Aubert:2004ns},
LHCb~\cite{Aaij:2011sn}, CMS~\cite{Chatrchyan:2013cld} and
BESIII~\cite{Ablikim:2013dyn} Collaborations. Its quantum number is
determined to be $I^{G}J^{PC}=0^{+}1^{++}$ by the LHCb
Collaboration~\cite{Aaij:2011sn}. Following the discovery of
$X(3872)$, a large number of charmoniumlike states have been
observed over the last decades, such as $X(3940)$~\cite{Abe:2004zs},
$X(4140)$~\cite{Aaltonen:2009tz}, $X(4160)$~\cite{Abe:2007sya},
$\psi(4260)$~\cite{Aubert:2005rm}, $\psi(4360)$~\cite{Aubert:2007zz},
$\psi(4660)$~\cite{Wang:2007ea}, and so on. These states have attracted
lots of attention from theorists. Various hadron configurations
including molecular state~\cite{Swanson:2003tb,Guo:2017jvc}, hybrid
meson~\cite{Zhu:2005hp,Esposito:2016itg},
tetraquark~\cite{Cui:2006mp,Park:2013fda}, etc. have been proposed to
explain their nature. A detailed review can be found in
Ref.~\cite{Chen:2016qju} and references therein.

Since the charmoniumlike states with normal quantum numbers have
similar masses compared to the normal charmonium, in order to
understand the nature of the exotic states, it is necessary to have
a better understanding of the normal charmonium spectroscopy. In
Ref.~\cite{Li:2009zu}, Li {\it et al.} investigated the spectrum of
higher charmonium with screened potential, and found that the vector
states $\psi(4008)$, $\psi(4260)$, $\psi(4320/4360)$, and $\psi(4660)$ might be
assigned as the $\psi(3S)$, $\psi(4S)$, $\psi(3D)$, and $\psi(6S)$
states, respectively, while $X(3940)$ and $X(4160)$ might be the
$\eta_{c}(3S)$ and $\chi_{c0}(3P)$ states. However, according to the
constituent quark model description by Segovia {\it et
al.}~\cite{Segovia:2013wma}, the mass of $\psi(4040)$, $\psi(4160)$,
$X(4360)$, $\psi(4415)$, $\psi(4630)$, and $\psi(4660)$ are compatible with
the $\psi(3S)$, $\psi(2D)$, $\psi(4S)$, $\psi(3D)$, $\psi(5S)$, and
$\psi(4D)$ states. Among the charmonium or charmoniumlike states,
the $1^{--}$ states are of special interest because they can be
easily produced in the $e^+e^-$ annihilation. In
Table~\ref{table:MASS}, we have listed the predicted masses of the
vector charmonium states from various models.

In addition, the decay properties of charmonium play a pivotal role
in revealing the nature of charmonium. From Table~\ref{table:MASS},
we see that the masses of these states are well above the allowed
two body open-charm decay threshold; thus the decay widths mainly
come from the strong decays. A widely used framework for the strong
decay is the quark pair creation (${^3P_0}$) model. In this model,
the $c\bar{c}$ pair in the initial charmonium regroups with a
$q\bar{q}$ pair created from the vacuum, which carries the vacuum
quantum number $J^{PC}=0^{++}$, and then decays into the outgoing
open-charm mesons. About forty years ago, Le~Yaouanc
{\it~et~al.}~\cite{LeYaouanc:1977fsz,LeYaouanc:1977gm} used this
model to study the open-charm strong decays of $\psi(4040)$ and
$\psi(4415)$. In 2005, Barnes {\it~et~al.} performed a systematic
study of the higher charmonium states just above $4.4~\text{GeV}$,
with the charmonium masses calculated in the Godfrey-Isgur (GI) model and a
nonrelativistic potential model~\cite{Barnes:2005pb}. In 2012,
Segovia {\it~et~al.}~\cite{Segovia:2013wma} studied the strong
decays of the vector charmonium states. For $\psi(3770)$,
$\psi(4040)$, $\psi(4160)$, and $\psi(4360)$ with $\psi(1D)$,
$\psi(3S)$, $\psi(2D)$, and $\psi(4S)$ assignments, the calculated
widths are compatible with the experimental values. While for the
$\psi(4415)$, $X(4640)$, and $\psi(4660)$ states, the difference between
theoretical and experimental values of the total widths is larger.
Recently, Gui {\it~et~al.}~\cite{Gui:2018rvv} studied the open-charm
strong decays of higher charmonium states up to the $6P$ multiplet
with their wave functions of charmonium states calculated in the
linear potential and screened potential quark models. Moreover, the
${^3P_0}$ model has also been used to study the strong decays of
bottomonium
states~\cite{Ferretti:2013vua,Godfrey:2015dia,Wang:2018rjg}.

\begin{table*}[ht]
\centering \caption{The predicted charmonium masses from various
models (in units of MeV).} \label{table:MASS}
\begin{tabular}{ccccccccccc}
\toprule[0.5pt]\toprule[0.5pt] \text{State}
~~~~~~&\text{QM}~\cite{Eichten:1979ms}
~~~~~~&\text{QM}~\cite{Godfrey:1985xj}
~~~~~~&\text{QM}~\cite{Segovia:2013wma}
~~~~~~&\text{SSE}/\text{EA}~\cite{Badalian:2008dv} ~~~~~~&\text{NR}/\text{GI}~\cite{Barnes:2005pb} ~~~~~~&\text{SP}~\cite{Li:2009zu} ~~~~~~&\text{LP}/\text{SP}~\cite{Deng:2016stx} \\
\midrule[0.5pt]
$\psi(3^3S_1)$ & 4225 & 4100 & 4097 & 4078/4096 & 4072/4100 & 4022 & 4078/4030 \\
$\psi(4^3S_1)$ & 4625 & 4450 & 4389 & 4398/4426 & 4406/4450 & 4273 & 4412/4281 \\
$\psi(5^3S_1)$ & $\cdots$ & $\cdots$ & 4614 & 4642/4672 & $\cdots$ & 4463 & 4711/4472 \\
$\psi(6^3S_1)$ & $\cdots$ & $\cdots$ & $\cdots$ & 4804/4828 & $\cdots$ & 4608 & $\cdots$ \\
$\psi(2^3D_1)$ & 4230 & 4190 & 4153 & 4156/4165 & 4142/4194 & 4089 & 4144/4095 \\
$\psi(3^3D_1)$ & $\cdots$ & 4520 & 4426 & 4464/4477 & $\cdots$ & 4317 & 4478/4336 \\
$\psi(4^3D_1)$ & $\cdots$ & $\cdots$ & 4641 & 4690/4707 & $\cdots$ & $\cdots$ & $\cdots$ \\
$\psi(5^3D_1)$ & $\cdots$ & $\cdots$ & $\cdots$ & 4840/4855 & $\cdots$ & $\cdots$ & $\cdots$ \\
\bottomrule[0.5pt]\bottomrule[0.5pt]
\end{tabular}
\end{table*}

Besides the two body decays, three body open flavor decay is also
important access to dig into the properties of charmonium and
bottomonium. In 2008, the Belle Collaboration~\cite{Pakhlova:2007fq}
first measured the exclusive cross section for
$e^+e^-{\to}D^0D^-\pi^+$ over the center-of-mass energy range
$(4.0-5.0)~\text{GeV}$ with the initial-state radiation (ISR) method and observed the decay
$\psi(4415){\to}D^0D^-\pi^+$. A detailed study found that the decay
is dominated by $\psi(4415){\to}D\bar{D}_{2}^{*}(2460)$ and
\begin{equation}
\frac{\mathcal{B}\left(\psi(4415){\to}D^0D^-\pi^+_{\text{nonresonant}}\right)}{\mathcal{B}\left(\psi(4415){\to}D\bar{D}_{2}^{*}(2460){\to}D^0D^-\pi^+\right)}
< 0.22
\end{equation}
at $90\%$ C.L. In 2009, they further measured the cross section of the
$e^+e^-{\to}D^0D^{*-}\pi^{+}+\text{c.c.}$ process and found no
evidence of $\psi(4260)$, $\psi(4360)$, $\psi(4415)$, $\psi(4630)$, or
$\psi(4660)$ with limited statistics~\cite{Pakhlova:2009jv}. Recently,
the BESIII Collaboration found two resonances in the
$e^+e^-{\to}D^0D^{*-}\pi^{+}$
process~\cite{Yuan:2018inv,Wang:2018osd,Ablikim:2018vxx}. The lower mass one is in
good agreement with the $Y(4220)$, and the other one might be
$\psi(4415)$. For the bottomonium state, the Belle Collaboration
also measured $\Upsilon(10860)$ decays into $B$
mesons~\cite{Drutskoy:2010an}. The measured fractions are
\begin{eqnarray}
 f\left(B\bar{B}\pi\right) = \left(0.0\pm1.1\pm0.3\right)\%, \\
 f\left(B\bar{B}^*\pi+B^*\bar{B}\pi\right) = \left(7.3_{-2.1}^{+2.3}\pm0.8\right)\%, \\
 f\left(B^*\bar{B}^*\pi\right) = \left(1.0_{-1.3}^{+1.4}\pm0.4\right)\%.
\end{eqnarray}
The measured three-body fractions are significantly larger than the
older predictions~\cite{Simonov:2008cr}.

\begin{figure}[ht]
\centering
\epsfxsize=5.0 cm
\epsfbox{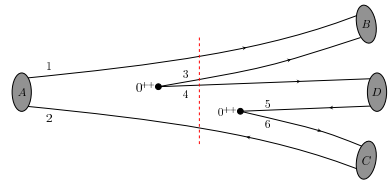}
\caption{The quarkonium ($A$) decays into three mesons ($B+C+D$).
The intermediate state is marked by a red dashed line.}
\label{fig:QPC_M3M}
\end{figure}

In Ref.~\cite{Xiao:2018iez}, we extended the ${^3P_0}$ model to
study the $\psi(4660){\to}\Lambda_{c}\bar{\Lambda}_{c}$ process with
two $q\bar{q}$ pairs created from the vacuum. In this paper, we
follow the extended ${^3P_0}$ model to study the three body open
flavor decays of higher charmonium and bottomonium states through a
different rearrangement (see Fig.~\ref{fig:QPC_M3M}). In the
framework of the extended ${^3P_0}$ model, we find that (i) the
$BB\pi$ and $B^*B^*\pi$ partial decay widths of the $\Upsilon(10860)$
state are consistent with the experiment. (For simplicity, we
abbreviate the $B\bar{B}\pi$, $B\bar{B}^*\pi+B^*\bar{B}\pi$, and
$B^*\bar{B}^*\pi$ to $BB\pi$, $BB^*\pi$, and $B^*B^*\pi$,
respectively. A similar abbreviation is also used for the charmonium
decays.) The $BB^*\pi$ partial decay width of the $\Upsilon(10860)$
state is smaller but very close to Belle's experiment. (ii) The
partial decay widths of the $DD^*\pi$ and $D^*D^*\pi$ modes can
reach up to several $\text{MeV}$ for the higher vector charmonium states. 
The three body open charm decay channels may be observed in the near
future.

For the singly heavy hadrons that contain one heavy quark, the heavy quark symmetry plays a very important role in the discussion of their properties such as masses and decay widths.
The theoretical framework is the well-known heavy quark effective theory (HQET).
With the help of HQET, the strong interaction between the heavy hadron multiplets may be described by one or two coupling constants, which can greatly simplify the discussion.
However, if there exist two heavy quarks, like the heavy quarkonium that we study here, the traditional heavy quark expansion does not work anymore because they may have different velocities.
%
%For the heavy quarkonium, a new effective field theory, NRQCD, was proposed.
%
In the present work, the strong decay widths for the heavy quarkonium states are restricted by many factors, such as the Clebsch-Gordan series, masses of the states and so on.
%
%So, the role of heavy quark symmetry is not well emphasized as explained above.
%

This paper is organized as follows. In Sec.~\ref{Sec:Model} the
${^3P_0}$ model and its extension are briefly introduced. The
numerical results are presented and discussed in
Sec.~\ref{Sec:Result}. Finally, a quick summary is given in
Sec.~\ref{Sec:Conclusion}.

%%%%%%%%%%%%%%%%%%%%%%%%%%%%
\section{The ${^3P_0}$ Model} \label{Sec:Model}
%%%%%%%%%%%%%%%%%%%%%%%%%%%%

The ${^3P_0}$ model is widely used to calculate the
Okubo-Zweig-Iizuka (OZI) allowed strong decays. It was first
proposed by Micu~\cite{Micu:1968mk} to study the strong decay
properties of the $P$-wave mesons. Le Yaouanc {\it et al.} of the
Orsay group further developed this model, and used it to study the
open flavor strong decays of
mesons~\cite{LeYaouanc:1972vsx,LeYaouanc:1977fsz,LeYaouanc:1977gm}
and baryons~\cite{LeYaouanc:1973ldf,LeYaouanc:1974cvx}. Since then,
this model has been widely used in the study of baryon strong
decays~\cite{Kalman:1988gg,Barnes:1996ff,Ackleh:1996yt,Barnes:2005pb,Close:2005se,Chen:2007xf,Godfrey:2015dia,Godfrey:2015dva}.
In the ${^3P_0}$ model, a light $q\bar{q}$ pair is created with the
vacuum quantum number $J^{PC}=0^{++}$ (hence ``the ${^3P_0}$
model''), and then rearranged with the quarks within the initial
meson to produce two final mesons. The decay matrix element can be
described by the interaction
Hamiltonian~\cite{Ackleh:1996yt,Barnes:2005pb,Close:2005se}
\begin{equation}\label{eqn:3P0:interaction}
H_{q\bar{q}} =
\gamma\sum_{f}2m_{f}\int\mathrm{d}^{3}{x}\bar{\psi}_{f}\psi_{f},
\end{equation}
where $m_{f}$ is the constituent quark mass of light flavor $f$, and $\psi_{f}$ is a
Dirac field of quark. $\gamma$ is a dimensionless constant standing
for the $q\bar{q}$ pair creation strength, which can be extracted by
fitting to data.

In Ref.~\cite{Xiao:2018iez}, we extended the ${^3P_0}$ model to
study the $\psi(4660){\to}\Lambda_{c}\bar{\Lambda}_{c}$ decay process,
which requires two light $q\bar{q}$ pairs to be created. Here, we go
a further step to study the higher heavy quarkonium decaying into
two heavy mesons plus a light meson, as shown in
Fig.~\ref{fig:QPC_M3M}. The corresponding helicity amplitude
$\mathcal{M}^{M_{J_A}M_{J_B}M_{J_C}M_{J_D}}$ is
\begin{eqnarray}\label{eqn:amplitude:definition}
\delta^{3}\left(\mathbf{p}_{A}-\mathbf{p}_{B}-\mathbf{p}_{C}-\mathbf{p}_{D}\right)\mathcal{M}^{M_{J_A}M_{J_B}M_{J_C}M_{J_D}}\nonumber\\
=\sum_{k}\frac{\Braket{BCD|H_{q\bar{q}}|k}\Braket{k|H_{q\bar{q}}|A}}{E_k-E_A},
\end{eqnarray}
where $\mathbf{p}_{i}$'s are the momenta of the hadrons; $\ket{k}$
stands for the intermediate state; $E_{A}$ and $E_{k}$ are energies
of the initial and intermediate states, respectively. We first
invoke the quark-hadron duality~\cite{Shifman:2000jv} to simplify
the calculation of the rhs of the
Eq.~(\ref{eqn:amplitude:definition}) at the quark level. More
precisely, we take $E_{k}-E_{A}$ as a constant, namely
$E_{k}-E_{A}\equiv2m_{q}$ since the intermediate state differs from
the initial state by a created $q\bar{q}$ pair at the quark
level~\cite{Xiao:2018iez} (see Fig.~\ref{fig:QPC_M3M}).
%To simplify the calculations, we take $E_{k}-E_{A}$ as a constant, namely $E_{k}-E_{A}\equiv2m_{q}$~\cite{Xiao:2018iez}.
%This approximation is reasonable because the intermediate state differs from the initial state by a created $q\bar{q}$ pair at the quark level.
%On the other hand, for charmonium state, the allowed intermediate states with the spin parity $J^{PC}=1^{--}$ are $D\bar{D}_1$, $D^*\bar{D}_1$, $D^*\bar{D}_0$, $D^*\bar{D}_2$, ${J/\psi}f_0(500)$, $h_c(1P)\pi$, $h_c(1P)\eta$, $\chi_{c0}(1P)\omega(782)$, $\chi_{c2}(1P)\omega(782)$ and tetraquark states~\cite{Maiani:2014aja} and so on.
%Their masses are about $\sim4.0~\text{GeV}$.
%Thus we have $E_{k}-E_{A}\sim2m_q$ for the higher mass charmonium states, especially the $\psi(4415)$ and $\psi(4660)$.
%Similarly, for bottomonium state, the allowed intermediate states with the quantum number $J^{PC}=1^{--}$ are $\eta_{b}(1S)h_1(1170)$, $\Upsilon(1S)f_0(980)$, $\chi_{b0}(1P)\omega(782)$, $\chi_{b1}(1P)\omega(782)$,  $\chi_{b2}(1P)\omega(782)$ and so on.
%Their masses are about $\sim10.5~\text{GeV}$.
%Thus we have $E_{k}-E_{A}\sim2m_q$ for the higher mass bottomonium states $\Upsilon(10860)$ and $\Upsilon(11020)$.
%Thus we simply assumed~\cite{Xiao:2018iez}
%\begin{equation}
%E_{k}-E_{A} \approx 2m_{q}
%\end{equation}
%where $m_q$ is the mass of the created quark.
Under the above approximation, we can rewrite the Eq.~(\ref{eqn:amplitude:definition}) as
\begin{eqnarray}\label{eqn:amplitude:definition:rewrite}
\delta^{3}\left(\mathbf{p}_{A}-\mathbf{p}_{B}-\mathbf{p}_{C}-\mathbf{p}_{D}\right)\mathcal{M}^{M_{J_A}M_{J_B}M_{J_C}M_{J_D}}\nonumber\\
= \frac{\Braket{BCD|H_{q\bar{q}}H_{q\bar{q}}|A}}{2m_{q}}.
\end{eqnarray}
The corresponding transition operator in the nonrelativistic limit
reads~\cite{Xiao:2018iez}
\begin{eqnarray}\label{eqn:3P0:transition-operator:2-pair}
T &=& \frac{9\gamma^2}{2m_{q}} \sum_{m,m'}
\Braket{1m;1-m|00}\Braket{1m';1-m'|00}\nonumber\\
&&
\times\int\mathrm{d}^3\mathbf{p}_3\mathrm{d}^3\mathbf{p}_4\mathrm{d}^3\mathbf{p}_5\mathrm{d}^3
\mathbf{p}_6\delta^3\left(\mathbf{p}_3+\mathbf{p}_4\right)\delta^3\left(\mathbf{p}_5+\mathbf{p}_6\right)\nonumber\\
&&\times\varphi^{34}_0\omega^{34}_0\chi^{34}_{1,-m}
\mathcal{Y}_1^m\left(\frac{\mathbf{p}_3-\mathbf{p}_4}{2}\right)a^{\dagger}_{3i}b^{\dagger}_{4j}\nonumber\\
&& \times\varphi^{56}_0\omega^{56}_0\chi^{56}_{1,-m'}
\mathcal{Y}_1^{m'}\left(\frac{\mathbf{p}_5-\mathbf{p}_6}{2}\right)a^{\dagger}_{5i'}b^{\dagger}_{6j'},
\end{eqnarray}
where $\mathbf{p}_i$ %($i$=3,~4,~5,~6)
is the momentum of the $i$th quark created from vacuum.
$\varphi_0=(u\bar{u}+d\bar{d}+s\bar{s})/\sqrt{3}$ and
$\omega_0=\delta_{ij}$ stand for the flavor and color singlets,
respectively. The solid harmonic polynomial
$\mathcal{Y}_1^{m(m')}(\mathbf{p})\equiv|\mathbf{p}|\mathcal{Y}^{m(m')}_1(\Omega_{\mathbf{p}})$
corresponds to the $P$-wave $q\bar{q}$ pair, and $\chi_{1,-m(m')}$
is the spin triplet state for the created $q\bar{q}$ pair.
$a^{\dagger}_{i}b^{\dagger}_{j}$ is the creation operator denoting
the $q\bar{q}$ pair creation in the vacuum.

We use the mock state~\cite{Hayne:1981zy} to define the meson ($A$)
\begin{eqnarray}\label{eqn:mock-meson}
&& 
\Ket{A\left(N_A~^{2S_A+1}L_A{J_AM_{J_A}}\right)\left(\mathbf{p}_A\right)}
\nonumber\\
&=&
\sqrt{2E_A}\varphi^{12}_A\omega^{12}_A\sum_{M_{L_A},M_{S_A}}
\Braket{L_AM_{L_A};S_AM_{S_A}|J_AM_{J_A}}
\nonumber\\
&\times&\int \mathrm{d}^3\mathbf{p}_1\mathrm{d}^3\mathbf{p}_2\delta^3(\mathbf{p}_1+\mathbf{p}_2-\mathbf{p}_A)\nonumber\\
&\times&
\Psi_{N_AL_AM_{L_A}}(\mathbf{p}_1,\mathbf{p}_2)\chi^{12}_{S_AM_{S_A}}
\Ket{q_1(\mathbf{p}_1)q_2(\mathbf{p}_2)}.
\end{eqnarray}
Here the $\mathbf{p}_{i}$ ($i=1,2$) is the momentum of quarks in
meson $A$. Then the helicity amplitude in the center-of-mass frame
can be written as
\begin{eqnarray}\label{eqn:amplitude}
&& \mathcal{M}^{M_{J_A}M_{J_B}M_{J_C}M_{J_D}}(A\to{BCD})\nonumber\\
&=&\frac{\gamma^2}{2m_q}\sqrt{16E_AE_BE_CE_D}\nonumber\\
&\times&\sum_{mm'}\sum_{M_{L_{A, B, C, D}},M_{S_{A,B,C,D}}}
\Braket{1m;1-m|00}
\Braket{1m';1-m'|00}
\nonumber\\
&\times&
\Braket{L_AM_{L_A}S_AM_{S_A}|J_AM_{J_A}}
\Braket{L_BM_{L_B}S_BM_{S_B}|J_BM_{J_B}}
\nonumber\\
&\times&
\Braket{L_CM_{L_C}S_CM_{S_C}|J_CM_{J_C}}
\Braket{L_DM_{L_D}S_DM_{S_D}|J_DM_{J_D}}
\nonumber\\
&\times&
\Braket{\chi_{S_BM_{S_B}}^{13}\chi_{S_CM_{S_C}}^{26}\chi_{S_DM_{S_D}}^{45}|\chi_{S_AM_{S_A}}^{12}\chi_{1,-m}^{34}\chi_{1,-m'}^{56}}
\nonumber\\
&\times&
\Braket{\varphi_{B}^{13}\varphi_{C}^{26}\varphi_{D}^{45}|\varphi_{A}^{12}\varphi_{0}^{34}\varphi_{0}^{56}}
\times
I_{M_{L_B}M_{L_C}M_{L_D}}^{M_{L_A}mm'}(\mathbf{p}),
\end{eqnarray}
where the factor $(-3)^2$ has been canceled by the color factor
\begin{equation}\label{eqn:color}
\Braket{\omega_{B}^{13}\omega_{C}^{26}\omega_{D}^{45}|\omega_{A}^{12}\omega_{0}^{34}\omega_{0}^{56}}
= \frac{1}{9}
\end{equation}
and $I_{M_{L_B},M_{L_C},M_{L_D}}^{M_{L_A},mm'}(\mathbf{p})$ is the
momentum-space integration and more detailed calculations are shown
in the Appendix~\ref{App:Sec:Integration}. Finally, the decay width $\Gamma$ reads
\begin{eqnarray}\label{eqn:gamma:3body}
\Gamma&=&\int_{0}^{\infty}dE_{B}dE_{C}\frac{\pi^3}{M_A}\frac{1}{2J_A+1}\nonumber\\
&\times&\sum_{M_{J_{A, B, C, D}}}
\left|\mathcal{M}^{M_{J_A}M_{J_B}M_{J_C}M_{J_D}}\right|^2.
\end{eqnarray}

Following the literature in this
field~\cite{LeYaouanc:1977fsz,Ackleh:1996yt,Barnes:1996ff,Barnes:2005pb,Close:2005se,Chen:2007xf,Godfrey:2015dia,Godfrey:2015dva},
we adopt the simple harmonic oscilator (SHO) wave function to
describe the momentum-space wave function of the meson
\begin{eqnarray}\label{eqn:SHO}
\psi_{nlm}^{\text{SHO}}\left(\mathbf{p}\right)
&=&\frac{(-1)^{n}(-i)^{l}}{\beta^{3/2}}
\sqrt{\frac{2n!}{\Gamma\left(n+l+3/2\right)}}
\left(\frac{p}{\beta}\right)^{l}\nonumber\\
&\times& \exp\left(-\frac{\mathbf{p}^2}{2\beta^2}\right)
L_{n}^{l+1/2}\left(\frac{p^2}{\beta^2}\right)
\mathcal{Y}_{l}^{m}\left(\Omega_{\mathbf{p}}\right),
\end{eqnarray}
where $L_{n}^{l+1/2}(p^2/\beta^2)$ is an associated Laguerre
polynomial.

In the present work, we set
$m_u=m_d=220~\text{MeV}$, $m_s=419~\text{MeV}$,
$m_c=1628~\text{MeV}$, and $m_b=4977~\text{MeV}$ for the constituent
quark masses~\cite{Godfrey:1985xj}. The masses of final state mesons
are listed in Table~\ref{table:mass&beta}. For simplicity, we ignore
the isospin breaking and obtain the meson masses by taking their
isospin averages.

\begin{table}[htbp]
    \centering \caption{Masses and harmonic oscillator strength
        $\beta$'s of final state mesons used in the decays (in units of
        MeV).} \label{table:mass&beta}
    \begin{tabular}{ccccccccccc}
        \toprule[0.5pt]\toprule[0.5pt]
        Meson ~~~~~& State ~~~~~& Mass~\cite{Tanabashi:2018oca} ~~~~~& $\beta$~\cite{Godfrey:2015dia,Godfrey:2015dva} \\
        \midrule[0.5pt]
        $\pi$ & ${^1S_0}$ & $138.0$ & $400$ \\
        %\quad$\pi^{0}$ && $134.9770\pm0.0005$ \\
        %\quad$\pi^{\pm}$ && $139.57061\pm0.00024$ \\
        $\rho$ & ${^3S_1}$ & $775.3$ & $400$ \\
        $\omega$ & ${^3S_1}$ & $782.6$ & $400$ \\
        %$K$ & ${^1S_0}$ & $495.644$ & $400$ \\
        %\quad$K^{0}$ && $497.611\pm0.013$ \\
        %\quad$K^{\pm}$ && $493.677\pm0.016$ \\
        %$K^{*}$ & ${^3S_1}$ & $893.66$ & $400$ \\
        %\quad$K^{*0}$ && $895.55\pm0.20$ \\
        %\quad$K^{*\pm}$ && $891.76\pm0.25$ \\
        $\eta$ & ${^1S_0}$ & $547.9$ & $400$ \\
        %\midrule[1pt]
        $D$ & ${^1S_0}$ & $1867.2$ & $600$ \\
        %\quad$D^{0}$ && $1864.83\pm0.05$ \\
        %\quad$D^{\pm}$ && $1869.65\pm0.05$ \\
        $D^{*}$ & ${^3S_1}$ & $2008.6$ & $520$ \\
        %\quad$D^{*0}$ && $2006.85\pm0.05$ \\
        %\quad$D^{*\pm}$ && $2010.26\pm0.05$ \\
        %\midrule[1pt]
        $D_{s}$ & ${^1S_0}$ & $1968.3$ & $650$ \\
        $D_{s}^{*}$ & ${^3S_1}$ & $2112.2$ & $560$ \\
        %\midrule[1pt]
        $B$ & ${^1S_0}$ & $5279.5$ & $580$ \\
        %\quad$B^{0}$ && $5279.63\pm0.15$ \\
        %\quad$B^{\pm}$ && $5279.32\pm0.14$ \\
        $B^{*}$ & ${^3S_1}$ & $5324.6$ & $540$ \\
        %\quad$B^{*0}$ && $$ \\
        %\quad$B^{*\pm}$ && $$ \\
        %\midrule[1pt]
        $B_{s}$ & ${^1S_0}$ & $5366.9$ & $640$ \\
        $B_{s}^{*}$ & ${^3S_1}$ & $5415.4$ & $600$ \\
        \bottomrule[0.5pt]\bottomrule[0.5pt]
    \end{tabular}
\end{table}

The harmonic oscillator strength $\beta$ of light mesons takes the
average value $400~\text{MeV}$~\cite{Godfrey:2015dva,Xiao:2018iez}.
The parameter $\beta$'s of heavy-light mesons (see Table~\ref{table:mass&beta}) are taken from
Refs.~\cite{Godfrey:2015dia,Godfrey:2015dva}, which are obtained by
the relation
\begin{equation}\label{eqn:relation:beta}
\int\mathrm{d}^3\mathbf{p}\,\left|\psi_{nlm}^{\text{SHO}}(\mathbf{p})\right|^2p^2=\int\mathrm{d}^3\mathbf{p}\,\left|\Phi(\mathbf{p})\right|^2p^2,
\end{equation}
where the lhs is the root-mean-square momentum of the SHO wave function, and the rhs is the root-mean-square momentum calculated through the GI model~\cite{Godfrey:2015dia,Godfrey:2015dva}.
%comparing the root-mean-square momentum of the SHO wave function to
%that of the wave functions calculated using the Godfrey-Isgur model
%(see Table~\ref{table:mass&beta}).
We use $\beta=500~\text{MeV}$ for
charmonium~\cite{Barnes:2003vb,Barnes:2005pb}. In
Ref.~\cite{Godfrey:2015dia}, Godfrey {\it et al.} showed that the
parameter $\beta$'s are $638$, $600$, and
$578~\text{MeV}$ for $\Upsilon(4{^3S_1})$,
$\Upsilon(5{^3S_1})$, and $\Upsilon(6{^3S_1})$,
respectively, thus we adopt the average value as $600~\text{MeV}$
for bottomonium states in this work.

For the $q\bar{q}$ pair creation strength, we use
$\gamma(c\bar{c})=6.95$ for charmonium decays, which is
$\sqrt{96\pi}$ times of that in
Refs.~\cite{Barnes:1996ff,Barnes:2005pb} due to a different
definition. However, in Ref.~\cite{Godfrey:2015dia} it is found that
this value underestimated the two body strong decay widths of
bottomonium, and the fitting of the open bottom decays of the
$\Upsilon$ sector gives $\gamma(b\bar{b})=10.42$. Here we adopt the
same value of $\gamma$ for $\Upsilon$ sector as in
Ref.~\cite{Godfrey:2015dia}. The uncertainty of $\gamma$ is about
$30\%$~\cite{Blundell:1996as,Close:2005se,Godfrey:2015dia,Godfrey:2015dva},
and the partial decay width is proportional to $\gamma^4$. Thus the uncertainty of our
results may be quite large.

\section{Numerical results}
\label{Sec:Result}

%\subsection{Parameters}
%\label{sec:parameter}

\subsection{Charmonium}
\label{sec:charmonium}

There are six charmoniulike states above the $DD\pi$ threshold,
$\psi(4040)$, $\psi(4160)$, $\psi(4260)$, $\psi(4360)$,
$\psi(4415)$, and $\psi(4660)$ with $J^{PC}=1^{--}$. These
states are of special interest since they can be easily produced
from the $e^{+}e^{-}$ annihilation. Note that the $\psi(4260)$ is
usually not considered to be a conventional charmonium
state~\cite{Gui:2018rvv}, so we do not discuss it in the following.
%We will discussed the three body decays in the following sections.

\subsubsection{$\psi(4360)$}%: $4S/3D$}
\label{sec:charmonium:4360}

The state $\psi(4360)$ was first observed by the $BABAR$ Collaboration in
the $e^{+}e^{-}\to\gamma_{\text{ISR}}\pi^+\pi^-\psi(2S)$
process~\cite{Aubert:2007zz}. Later, the Belle Collaboration
confirmed this state in the same process with a statistical
significance of more than $8\sigma$~\cite{Wang:2007ea}. The average
values of mass and width listed in PDG are $M=4368\pm13~\text{MeV}$
and
$\Gamma_{\text{tot.}}=96\pm7~\text{MeV}$~\cite{Tanabashi:2018oca}.
An interesting feature is that only the $\psi(2S)\pi^+\pi^-$ [and
possibly $\psi_2(3823)\pi^+\pi^-$] decay mode(s) was observed, while
the open charm decay modes are still
missing~\cite{Aubert:2007zz,Wang:2007ea,Ablikim:2015dlj}.

$\psi(4360)$ was interpreted to be a $3{^3D_1}$ state in the
nonrelativistic screened potential model~\cite{Li:2009zu}. Ding {\it
et al.} also interpreted $\psi(4360)$ as a $3{^3D_1}$ charmonium by
evaluating its $e^+e^-$ leptonic widths, E1 transitions, M1
transitions and the open flavor strong decays in the flux tube
model. However, the possibility of the $4{^3S_1}$ assignment cannot
be ruled out~\cite{Segovia:2013wma}. As the possible assignments of
$\psi(4360)$, it is crucial to study the decay properties of the
$\psi(4{^3S_1})$ and $\psi(3{^3D_1})$. The
theoretical predictions are listed in
Table~\ref{table:width:cc:4360:DDpi}.

\begin{table}[htbp]
\centering \caption{The partial decay widths (in MeV) of the vector
charmonium with a mass of $4368~\text{MeV}$.}
\label{table:width:cc:4360:DDpi}
\begin{tabular}{lccccccccccc}
\toprule[0.5pt]\toprule[0.5pt]
State ~~~~~~~&$\psi\left(4{^3S_1}\right)$ ~~~~~~~&$\psi\left(3{^3D_1}\right)$ \\
\midrule[0.5pt]
$\Gamma_{DD\pi}$ ~~~~~~~& $0.27$ ~~~~~~~& $0.14$ \\
$\Gamma_{DD^{*}\pi}$ ~~~~~~~& $1.40$ ~~~~~~~& $1.21$ \\
$\Gamma_{D^{*}D^{*}\pi}$ ~~~~~~~& $0.60$ ~~~~~~~& $0.25$ \\
$\Gamma_{DD\eta}$ ~~~~~~~& $0.6\text{ keV}$ ~~~~~~~& $0.3\text{ keV}$ \\
\bottomrule[0.5pt]\bottomrule[0.5pt]
\end{tabular}
\end{table}

From Table~\ref{table:width:cc:4360:DDpi}, the dominant three
body decay mode for both $\psi(4{^3S_1})$ and
$\psi(3{^3D_1})$ is $D\bar{D}^{*}\pi$ with a mass of
$M$=4368 MeV, and the predicted partial decay widths are
\begin{eqnarray}
& \Gamma[\psi\left(4{^3S_1}\right){\to}DD^*\pi] \sim
1.40~\text{MeV},
\end{eqnarray}
and
\begin{eqnarray}
& \Gamma[\psi\left(3{^3D_1}\right){\to}DD^*\pi] \sim
1.21~\text{MeV}.
\end{eqnarray}
Combing the measured width of $\psi(4360)$, we further get the
branching ratios
\begin{eqnarray}
& \mathcal{B}[\psi\left(4{^3S_1}\right){\to}DD^*\pi] \sim 1.5\%,\\
& \mathcal{B}[\psi\left(3{^3D_1}\right){\to}DD^*\pi] \sim 1.3\%.
\end{eqnarray}
The sizeable branching ratios indicates that this state has a good
potential to be observed in the $DD^*\pi$ decay channel if it indeed
turns out to be either the state $\psi(4{^3S_1})$ or
$\psi(3{^3D_1})$.

Meanwhile, the partial decay widths of $DD\pi$ and $D^*D^*\pi$ are
sizable for the two assignments. If $\psi(4360)$ is the $4{^3S_1}$
state, we predict
\begin{equation}
\Gamma(DD\pi) : \Gamma(DD^*\pi) : \Gamma(D^*D^*\pi) \sim
1.0:5.1:2.2,
\end{equation}
while the $3{^3D_1}$ assignment gives
\begin{equation}
\Gamma(DD\pi) : \Gamma(DD^*\pi) : \Gamma(D^*D^*\pi) \sim
1.0:8.1:1.7.
\end{equation}

The $DD\eta$ decay mode is also available kinetically. However, our
calculation shows that its width
[$\mathcal{O}\left(0.1~\text{keV}\right)$] is too small to be
observed because of its tiny phase space.

Unfortunately the three body decay properties of the two assignments
$\psi(4{^3S_1})$ and $\psi(3{^3D_1})$ are very
similar, which cannot be used to distinguish these two states in
future experiments.

\subsubsection{$\psi(4415)$}%: $4{^3S_1}/3D$}
\label{sec:charmonium:4415}

The $\psi(4415)$ state was discovered by SLAC and LBL in $e^+e^-$
annihilation~\cite{Siegrist:1976br}. Later, it was confirmed by the DASP
Collaboration~\cite{Brandelik:1978ei}. Its mass and width are
$(4421\pm4)$ and
$(62\pm20)~\text{MeV}$~\cite{Tanabashi:2018oca}, respectively. This
state is the unique vector charmonium with experimental data of
three body decays. The present study of the state $\psi(4415)$ can
not only provide an important test of our model but also let us
obtain more information about the nature of $\psi(4415)$.

In Ref.~\cite{LeYaouanc:1977gm}, Le Yaouanc {\it et al.} used the
${^3P_0}$ model to calculate its open flavor decay and assigned it
to be the $4{^3S_1}$ state. Later, Barnes {\it et al.} confirmed this
assignment by comparing the mass spectrum from GI model calculation.
They calculated all ten open-charmed decay widths of $\psi(4415)$
using the ${^3P_0}$ model, and found that the total widths and the
decay patterns were consistent with
experiments~\cite{Barnes:2005pb}. Moreover, they predicted that
$DD_1$ and $DD_2^*$ were the major decay modes of $\psi(4415)$, and
the latter prediction was confirmed by Belle
Collaboration~\cite{Pakhlova:2007fq}. Thus it is essential to study
the three body decay properties of $\psi(4{^3S_1})$.

\begin{table}[htbp]
\centering \caption{The partial decay widths (in MeV) of the vector
charmonium with a mass of $4421~\text{MeV}$.}
\label{table:width:cc:4415:DDpi}
\begin{tabular}{lccccccccccc}
\toprule[0.5pt]\toprule[0.5pt]
State ~~~~~&$\psi\left(4{^3S_1}\right)$ ~~~~~&$\psi\left(5{^3S_1}\right)$ ~~~~~&$\psi\left(3{^3D_1}\right)$ \\
\midrule[0.5pt]
$\Gamma_{DD\pi}$ ~~~~~& $0.38$ ~~~~~& $0.11$ ~~~~~& $0.21$ \\
$\Gamma_{DD^{*}\pi}$ ~~~~~& $2.01$ ~~~~~& $0.96$ ~~~~~& $1.84$ \\
$\Gamma_{D^{*}D^{*}\pi}$ ~~~~~& $1.07$ ~~~~~& $0.59$ ~~~~~& $0.52$ \\
$\Gamma_{DD\eta}$ ~~~~~& $5.4\text{ keV}$ ~~~~~& $1.7\text{ keV}$ ~~~~~& $2.9\text{ keV}$ \\
\bottomrule[0.5pt]\bottomrule[0.5pt]
\end{tabular}
\end{table}

\begin{figure*}[ht]
\centering \epsfxsize=15.0 cm \epsfbox{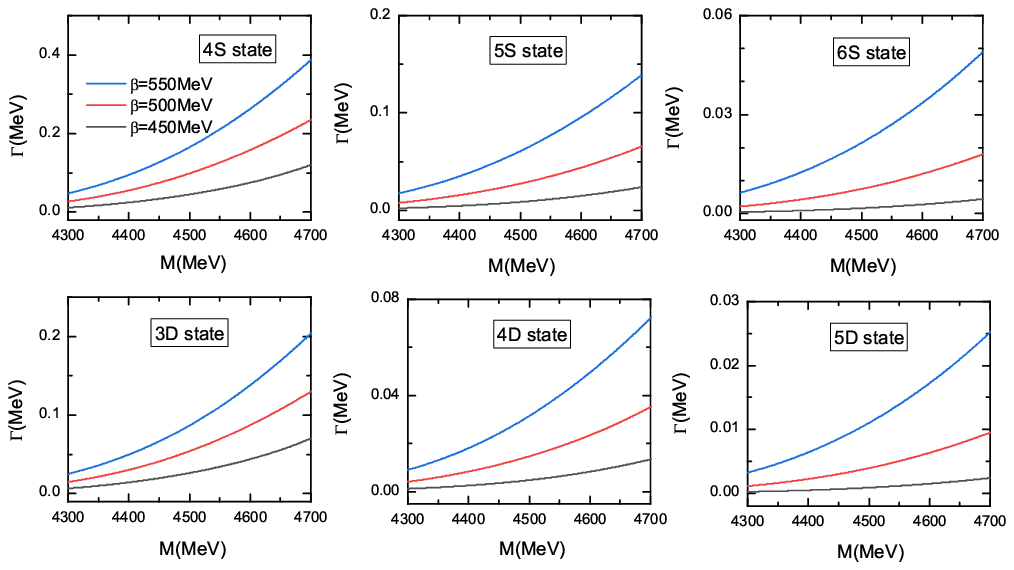} \caption{The
variation of the $D^+D^-\pi^0$ partial decay width with the mass of
the $D$-wave vector charmonium. Note that
$\Gamma_{D^{+}D^{-}\pi^{0}}=\frac{1}{6}\Gamma_{DD\pi}$ since we have
ignored the isospin breaking. The blue, red, and black lines
correspond to the predictions with different values of the harmonic
oscillator strength $\beta=450,~500,~\text{and}~550~\text{MeV}$,
respectively.}\label{fig:DDpi}
\end{figure*}

\begin{figure*}[ht]
\centering \epsfxsize=15.0 cm \epsfbox{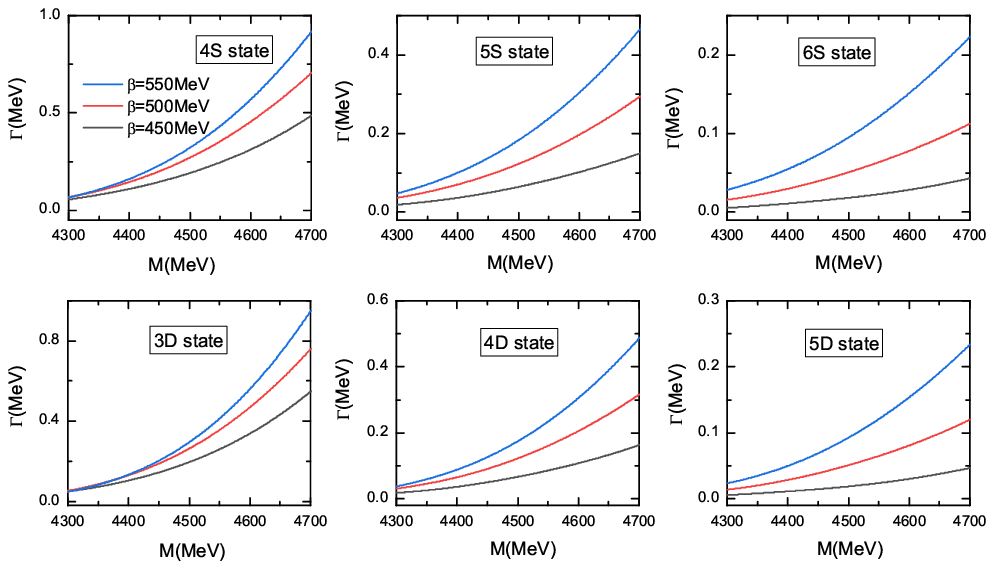} \caption{The
variation of the $D^+D^{*-}\pi^0$ partial decay width with the mass
of the $D$-wave vector charmonium. Note that
$\Gamma_{D^{+}D^{*-}\pi^{0}}=\frac{1}{12}\Gamma_{DD^*\pi}$ since we
have ignored the isospin breaking. The blue, red, and black lines
correspond to the predictions with different values of the harmonic
oscillator strength $\beta=450,~500,~\text{and}~550~\text{MeV}$,
respectively.}\label{fig:DDstarpi}
\end{figure*}

\begin{figure*}[ht]
\centering \epsfxsize=15.0 cm \epsfbox{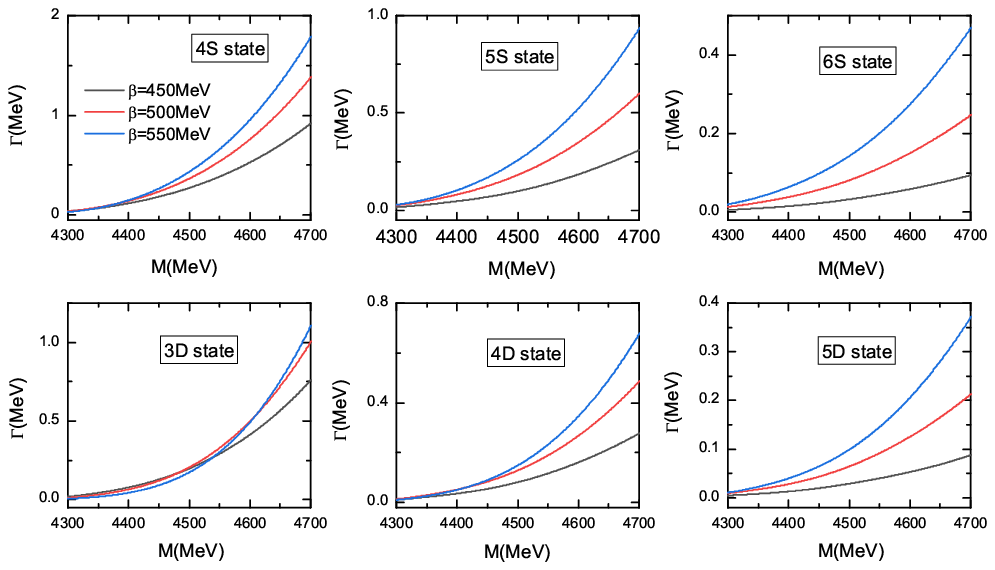}
\caption{The variation of the $D^{*+}D^{*-}\pi^0$ partial decay
width with the mass of the $D$-wave vector charmonium. Note that
$\Gamma_{D^{*+}D^{*-}\pi^{0}}=\frac{1}{6}\Gamma_{D^*D^*\pi}$ since
we have ignored the isospin breaking. The blue, red, and black lines
correspond to the predictions with different values of the harmonic
oscillator strength $\beta=450,~500,~\text{and}~550~\text{MeV}$,
respectively.}\label{fig:DstarDstarpi}
\end{figure*}

Fixing the mass of $\psi(4{^3S_1})$ at
$M=4421~\text{MeV}$, we calculated its partial decay widths and
listed them in Table~\ref{table:width:cc:4415:DDpi}. According to
our calculation, its three body strong decay is governed by the
$DD^*\pi$ channel with the branching ratio
\begin{equation}
\mathcal{B}[\psi(4{^3S_1}){\to}DD^*\pi] \sim 3.2\%,
\end{equation}
which is less than the upper limit ($<11\%$) listed in
PDG~\cite{Tanabashi:2018oca}.

The role of the $D^*D^*\pi$ channel is also important in the decays.
The predicted partial width ratio between $D^*D^*\pi$ and $DD^*\pi$
is
\begin{equation}
\frac{\Gamma[\psi(4{^3S_1}){\to}D^*D^*\pi]}{\Gamma[\psi(4{^3S_1}){\to}DD^*\pi]}
\sim 0.5.
\end{equation}

Meanwhile, the partial decay width of the $DD\pi$ mode is predicted
to be
\begin{equation}
\Gamma[\psi(4{^3S_1}){\to}DD\pi] \sim 0.38~\text{MeV}
\end{equation}
with the branching ratio
\begin{equation}
\mathcal{B}[\psi(4{^3S_1}){\to}DD\pi] \sim 0.6\%.
\end{equation}
This value is also less than the upper limit ($2.2\%$) obtained by
the Belle Collaboration~\cite{Pakhlova:2007fq}. To further confirm
the nature of $\psi(4415)$ and test our results, more precise
experimental data are badly needed.

Besides the $4{^3S_1}$ assignment, there are other interpretations
of $\psi(4415)$. In Ref.~~\cite{Li:2009zu}, Li {\it~et~al.} proposed
that the mass of $\psi(4415)$ was compatible with $5{^3S_1}$ rather
than $4{^3S_1}$ in the nonrelativistic screened potential model.
However, Segovia {\it~et~al.} suggested that $\psi(4415)$ could be a
$3{^3D_1}$ state~\cite{Segovia:2013wma}. In the present work, we
also calculate the partial decay widths of
$\psi(5{^3S_1})$ and $\psi(3{^3D_1})$ with the
mass of $4421~\text{MeV}$. The predictions are collected in
Table~\ref{table:width:cc:4415:DDpi}.

Fixing the masses at $M=4421$ MeV, the dominant three body decay
mode of $\psi(5{^3S_1})$ and $\psi(3{^3D_1})$
is $DD^*\pi$. The partial decay widths are
\begin{equation}
\Gamma[\psi(5{^3S_1}){\to}DD^*\pi] \sim 0.96~\text{MeV}
\end{equation}
and
\begin{equation}
\Gamma[\psi(3{^3D_1}){\to}DD^*\pi] \sim 1.84~\text{MeV}.
\end{equation}
The above decay widths are large enough to be observed in future
experiments. Moreover,
\begin{equation}
\frac{\Gamma[\psi(3{^3D_1}){\to}DD^*\pi]}{\Gamma[\psi(5{^3S_1}){\to}DD^*\pi]}
\sim 1.9,
\end{equation}
which indicates that the $DD^*\pi$ branching ratio of $\psi(3{^3D_1})$ 
is larger that of $\psi(5{^3S_1})$.

In addition, considering the uncertainty of the predicted masses in
various models, we plot the partial decay widths of the $4{^3S_1}$,
$5{^3S_1}$, and $3{^3D_1}$ $c\bar{c}$ states as the functions of the
mass in the range of $M=(4300-4700)~\text{MeV}$ in
Figs.~\ref{fig:DDpi}--\ref{fig:DstarDstarpi}.

\subsubsection{$\psi(4660)$}%: 5S/4D}
\label{sec:charmonium:4660}

In 2007, the Belle Collaboration reported an enhancement $\psi(4660)$
when they measured the cross section of
the $e^+e^-\to\pi^+\pi^-\psi(2S)$ process~\cite{Wang:2007ea}. Later, the
$BABAR$ Collaboration confirmed the existence of the $\psi(4660)$ state
in the same process~\cite{Lees:2012pv}. The mass and width of the
$\psi(4660)$ are $(4643\pm9)$ and $(72\pm11)~\text{MeV}$,
respectively. According to various quark model calculations, there
are six excited vector charmonium states around 4.6 GeV (see
Table~\ref{table:MASS}), namely $\psi(4S)$, $\psi(5S)$, $\psi(6S)$,
$\psi(3D)$, $\psi(4D)$, and $\psi(5D)$. In Ref.~\cite{Xiao:2018iez},
we have studied the $\Lambda_{c}\bar{\Lambda}_{c}$ partial decay
widths of these states. Here we discuss their three body
decays.

According to the quark model calculation, the mass of
$\psi(5{^3S_1})$ is very close to $\psi(4660)$. Ding {\it et
al.} also suggested that the $\psi(4660)$ is a $5{^3S_1}$ charmonium
after studying its $e^+e^-$ leptonic widths, E1 transitions, M1
transitions and the open flavor strong decays in the flux-tube
model~\cite{Ding:2007rg}. As a possible assignment, we first study
the decay property of the $\psi(5{^3S_1})$ and list the
corresponding results in Table~\ref{table:width:cc:4660:DDpi}.

\begin{table*}[htbp]
\centering \caption{The partial decay width (MeV) of the vector
charmonium with a mass of $4643~\text{MeV}$.}
\label{table:width:cc:4660:DDpi}
\begin{tabular}{lcccccc}
\toprule[0.5pt]\toprule[0.5pt]
State ~~~~~~~&$\psi\left(4{^3S_1}\right)$ ~~~~~~~&$\psi\left(5{^3S_1}\right)$ ~~~~~~~&$\psi\left(6{^3S_1}\right)$ ~~~~~~~&$\psi\left(3{^3D_1}\right)$ ~~~~~~~&$\psi\left(4{^3D_1}\right)$ ~~~~~~~&$\psi\left(5{^3D_1}\right)$ \\
\midrule[0.5pt]
$\Gamma_{DD\pi}$ ~~~~~~~& $1.14$ ~~~~~~~& $0.31$ ~~~~~~~& $0.09$ ~~~~~~~& $0.63$ ~~~~~~~& $0.17$ ~~~~~~~& $0.05$ \\
$\Gamma_{DD^*\pi}$ ~~~~~~~& $6.65$ ~~~~~~~& $2.83$ ~~~~~~~& $1.10$ ~~~~~~~& $6.99$ ~~~~~~~& $2.99$ ~~~~~~~& $1.16$ \\
$\Gamma_{D^*D^*\pi}$ ~~~~~~~& $5.97$ ~~~~~~~& $2.68$ ~~~~~~~& $1.13$ ~~~~~~~& $4.12$ ~~~~~~~& $2.11$ ~~~~~~~& $0.96$ \\
$\Gamma_{DD\rho}$ ~~~~~~~& $0.85$ ~~~~~~~& $0.41$ ~~~~~~~& $0.16$ ~~~~~~~& $1.86$ ~~~~~~~& $0.64$ ~~~~~~~& $0.22$ \\
$\Gamma_{DD\omega}$ ~~~~~~~& $0.24$ ~~~~~~~& $0.12$ ~~~~~~~& $0.05$ ~~~~~~~& $0.59$ ~~~~~~~& $0.20$ ~~~~~~~& $0.07$ \\
$\Gamma_{DD\eta}$ ~~~~~~~& $53.2~\text{keV}$ ~~~~~~~&
$15.3~\text{keV}$ ~~~~~~~& $4.2~\text{keV}$ ~~~~~~~&
$29.1~\text{keV}$ ~~~~~~~& $8.2~\text{keV}$ ~~~~~~~& $2.2~\text{keV}$ \\
$\Gamma_{DD^*\eta}$ ~~~~~~~& $0.25$ ~~~~~~~& $0.12$ ~~~~~~~& $0.05$ ~~~~~~~& $0.20$ ~~~~~~~& $0.11$ ~~~~~~~& $0.05$ \\
$\Gamma_{D^*D^*\eta}$ ~~~~~~~& $58.2~\text{keV}$ ~~~~~~~&
$38.7~\text{keV}$ ~~~~~~~& $19.1~\text{keV}$ ~~~~~~~&
$8.5~\text{keV}$ ~~~~~~~& $9.9~\text{keV}$ ~~~~~~~& $7.5~\text{keV}$ \\
$\Gamma_{D_{s}D_{s}\eta}$ ~~~~~~~& $3.0~\text{keV}$ ~~~~~~~&
$0.8~\text{keV}$ ~~~~~~~& $0.2~\text{keV}$ ~~~~~~~&
$1.6~\text{keV}$ ~~~~~~~& $0.4~\text{keV}$ ~~~~~~~& $0.1~\text{keV}$ \\
$\Gamma_{D_{s}D_{s}^*\eta}$ ~~~~~~~& $1.9~\text{keV}$ ~~~~~~~&
$1.3~\text{keV}$ ~~~~~~~& $0.6~\text{keV}$ ~~~~~~~&
$12~\text{eV}$ ~~~~~~~& $11~\text{eV}$ ~~~~~~~& $7~\text{eV}$ \\
\bottomrule[0.5pt]\bottomrule[0.5pt]
\end{tabular}
\end{table*}

From the table, we find that the partial decay widths of $DD^{*}\pi$
and $D^*D^*\pi$ modes are quite large, which read
\begin{eqnarray}
& \mathcal{B}[\psi(5{^3S_1}){\to}DD^*\pi] \sim 3.9\%
\end{eqnarray}
and
\begin{equation}
\mathcal{B}[\psi(5{^3S_1}){\to}D^*D^*\pi] \sim 3.7\%,
\end{equation}
respectively. The values are large enough to be observed in
experiment. Meanwhile, the partial decay width of
$\Gamma[\psi(5{^3S_1}){\to}DD\pi]$ is considerable. The
partial decay width ratio is
\begin{eqnarray}
\frac{\Gamma[\psi(5{^3S_1}){\to}DD\pi]}{\Gamma[\psi(5{^3S_1}){\to}DD^*\pi]}\sim
0.1.
\end{eqnarray}

Besides $\psi(5{^3S_1})$, the possibility that $\psi(4660)$
is a $\psi(4{^3S_1})$ or $\psi(6{^3S_1})$
state cannot be excluded completely. Thus we also calculate the
partial decay widths of the $\psi(4{^3S_1})$ and $\psi(6{^3S_1})$
states. Similarly, we fix the mass of $\psi(4{^3S_1})$ and
$\psi(6{^3S_1})$ at $M=4643~\text{MeV}$, and collect their partial
decay widths in Table~\ref{table:width:cc:4660:DDpi}.

As listed in Table~\ref{table:width:cc:4660:DDpi}, we obtain that
the partial decay widths of $DD^{*}\pi$ and $D^*D^*\pi$ for
$\psi(4{^3S_1})$ are the largest compared to those for
$\psi(5{^3S_1})$ and $\psi(6{^3S_1})$. The predicted branching
ratios are
\begin{eqnarray}
& \mathcal{B}[\psi(4{^3S_1}){\to}DD^*\pi] \sim 9.2\%, \nonumber\\
& \mathcal{B}[\psi(4{^3S_1}){\to}D^*D^*\pi] \sim 8.3\%.
\end{eqnarray}
However, the $\psi(6{^3S_1})$ state gives the smallest
branching ratios, which are
\begin{eqnarray}
& \mathcal{B}[\psi(6{^3S_1}){\to}DD^*\pi] \sim 1.5\%, \nonumber\\
& \mathcal{B}[\psi(6{^3S_1}){\to}D^*D^*\pi] \sim 1.6\%.
\end{eqnarray}

Furthermore, we study the decay properties of the
$\psi(3{^3D_1})$, $\psi(4{^3D_1})$, and
$\psi(5{^3D_1})$ states, and list their decay properties
in Table~\ref{table:width:cc:4660:DDpi} as well. Combined with the
total width of $\psi(4660)$, we obtain the branching ratios of the
$\psi(3{^3D_1})$, $\psi(4{^3D_1})$, and
$\psi(5{^3D_1})$ states as follows:
\begin{eqnarray}
& \mathcal{B}[\psi(3{^3D_1}){\to}DD\pi] \sim 0.6\%, \\
& \mathcal{B}[\psi(3{^3D_1}){\to}DD^*\pi] \sim 9.7\%, \\
& \mathcal{B}[\psi(3{^3D_1}){\to}D^*D^*\pi] \sim 5.7\%;
\end{eqnarray}
\begin{eqnarray}
& \mathcal{B}[\psi(4{^3D_1}){\to}DD\pi] \sim 0.2\%, \\
& \mathcal{B}[\psi(4{^3D_1}){\to}DD^*\pi] \sim 4.2\%, \\
& \mathcal{B}[\psi(4{^3D_1}){\to}D^*D^*\pi] \sim 2.9\%;
\end{eqnarray}
\begin{eqnarray}
& \mathcal{B}[\psi(5{^3D_1}){\to}DD\pi] \sim 0.06\%, \\
& \mathcal{B}[\psi(5{^3D_1}){\to}DD^*\pi] \sim 1.6\%, \\
& \mathcal{B}[\psi(5{^3D_1}){\to}D^*D^*\pi] \sim 1.3\%.
\end{eqnarray}
These branching ratios are comparable to those of the $S$-wave
states. If $\psi(4660)$ is a $D$-wave state, it is possible to be
observed in the $DD^*\pi$ and $D^*D^*\pi$ channels as well.

In addition to the $DD\pi$, $DD^*\pi$, and $D^*D^*\pi$ channels,
$\psi(4660)$ can also decay into $DD\rho$, $DD\omega$, $DD\eta$,
$DD^*\eta$, $D^*D^*\eta$, $D_sD_s\eta$, and $D_sD_s^*\eta$ channels.
In the same way, we fix the mass of the states $\psi(4S)$,
$\psi(5S)$, $\psi(6S)$, $\psi(3D)$, $\psi(4D)$, and $\psi(5D)$ at
$M=4643~\text{MeV}$, and calculate their widths of decaying into
these channels. The results are collected in
Table~\ref{table:width:cc:4660:DDpi}. The partial decay widths of
these channels are relatively smaller. Among them, the partial decay
widths of the $DD\rho$, $DD\omega$, and $DD^*\eta$ modes are around
several tenths of MeV. If $\psi(4660)$ is one of the above states, it
is still possible to observe these channels.

The mass spectrum predicted by various quark models bears a large
uncertainty, and may have an effect on the partial decay widths. To
investigate this effect, we vary the mass of the states $\psi(4S)$,
$\psi(5S)$, $\psi(6S)$, $\psi(3D)$, $\psi(4D)$, and $\psi(5D)$ from
$4300$ to $4700~\text{MeV}$, and calculate their
corresponding decay widths. Here, we just plot the results for the
$DD\pi$, $DD^*\pi$, and $D^*D^*\pi$ channels in
Figs.~\ref{fig:DDpi}--\ref{fig:DstarDstarpi}, and omit the
theoretical predictions of other channels since their decay widths
are relatively smaller.

\subsubsection{$\psi(4040)$ and $\psi(4160)$}%: $3{^3S_1}$}
\label{sec:charmonium:4040-4160}

There are two remaining states with $J^{PC}=1^{--}$, namely the $\psi(4040)$ and $\psi(4160)$ state.
The $\psi(4040)$ state is commonly believed to be the $3{^3S_1}$
$c\bar{c}$ state~\cite{Barnes:2005pb}. Its average mass and width
are $M=4039\pm1~\text{MeV}$ and
$\Gamma_{\text{tot.}}=80\pm10~\text{MeV}$~\cite{Tanabashi:2018oca},
respectively. This is the first $1^{--}$ charmonium above the
$D\bar{D}\pi$ threshold.
The mass of $\psi(4160)$ is
$4191\pm5~\text{MeV}$~\cite{Tanabashi:2018oca}, which is about 150
MeV heavier than that of $\psi(4040)$. According to the mass
predictions in the quark model~\cite{Barnes:2005pb}, this state is
suggested to be the $2{^3D_1}$ $c\bar{c}$ state. Its two body open
charm decays have been studied by many authors, which also support
this assignment~\cite{LeYaouanc:1977fsz,Barnes:2005pb,Gui:2018rvv}.

\begin{table}[htbp]\centering
\caption{The $D^{(*)}D^{(*)}\pi$ partial decay width (in keV) for
the two established charmonium states $\psi(4040)$ and
$\psi(4160)$.} \label{table:width:cc:4040+4160:DDpi}
\begin{tabular}{cccccccccc}
\toprule[0.5pt]\toprule[0.5pt]
Meson ~~& State ~~& Mode ~~&
$\Gamma_{\text{th}}^{\beta=450~\text{MeV}}$ ~&
$\Gamma_{\text{th}}^{\beta=500~\text{MeV}}$ ~&
$\Gamma_{\text{th}}^{\beta=550~\text{MeV}}$ \\
\midrule[0.5pt]
$\psi(4040)$ & $3{^3S_1}$ &$D\bar{D}\pi$ & $14.7$ & $20.9$ & $24.0$ \\
&& $DD^{*}\pi$ & $15.1$ & $3.2$ & $0.5$ \\
$\psi(4160)$ & $2{^3D_1}$ &$DD\pi$ & $80.4$ & $119.8$ & $142.8$ \\
&& $DD^{*}\pi$ & $188.3$ & $132.6$ & $134.3$ \\
&& $D^{*}D^{*}\pi$ & $0.4$ & $0.1$ & $0.2$ \\
\bottomrule[0.5pt]\bottomrule[0.5pt]
\end{tabular}
\end{table}

We present our results in Table~\ref{table:width:cc:4040+4160:DDpi}.
According to our calculation, the partial decay width is
\begin{equation}
\Gamma[\psi(4040){\to}D{D}\pi] \sim 20.9\text{ keV},
\end{equation}
which is quite small compared to the total decay width of
$\psi(4040)$. More precisely, the branching ratio is
\begin{equation}
\mathcal{B}[\psi(4040){\to}DD\pi] \sim 2.6\times10^{-4}.
\end{equation}
This ratio is smaller than that of the hidden charm decay modes of
$\psi(4040)$ by 1 order. Because of the narrow partial decay width,
the $DD\pi$ decay mode might be not easy to be observed.

The the $DD^{*}\pi$ mode is also available for $\psi(4040)$. Since
$DD^*\pi$ mode has little phase space, the partial width of
$\psi(4040)$ decaying into $DD^*\pi$ is about one magnitude smaller
than the $D{D}\pi$ partial width. The partial decay width ratio
is
\begin{equation}
\frac{\Gamma[\psi(4040){\to}DD^*\pi]}{\Gamma[\psi(4040){\to}DD\pi]}
\sim 0.2.
\end{equation}

%\subsubsection{$\psi(4160)$}%: $2{^3D_1}$}
%\label{sec:charmonium:4160}

We also analyze the three body decay properties of $\psi(4160)$ as the
$2{^3D_1}$ $c\bar{c}$ state, and collect its partial strong decay
widths in Table ~\ref{table:width:cc:4040+4160:DDpi}. We obtain the
partial decay widths
\begin{equation}
\Gamma[\psi(4160){\to}DD\pi] \sim 119.8~\text{keV}
\end{equation}
and
\begin{equation}
\Gamma[\psi(4160){\to}DD^*\pi] \sim 132.6\text{ keV}.
\end{equation}
The values are much bigger than the corresponding one of the
$\psi(4040)$. These widths seem not large compared to its total
width ($\Gamma_{\text{tot}.}=70\pm10$ MeV), but it is enough to be
observed in those decay channels in experiments. Moreover, the
branching ratios are predicted to be
\begin{equation}
\mathcal{B}[\psi(4160){\to}DD\pi] \sim 1.7\times10^{-3}
\end{equation}
and
\begin{equation}
\mathcal{B}[\psi(4160){\to}DD^*\pi] \sim 1.9\times10^{-3},
\end{equation}
which are comparable to the upper limit of hidden charm decays of
$\psi(4160)$. The partial decay width of $D^*D^*\pi$ mode is
\begin{equation}
\Gamma[\psi(4160){\to}D^*D^*\pi] \sim 0.1\text{ keV}.
\end{equation}
This value is small and hard to be searched for at present.

The results of $\psi(4040)$ and $\psi(4160)$ may have large
uncertainties due to their lower masses. At the hadron level, the
allowed intermediate states with the spin parity $J^{PC}=1^{--}$ are
$D\bar{D}_1$, $D^*\bar{D}_1$, $D^*\bar{D}_0$, $D^*\bar{D}_2$,
${J/\psi}f_0(500)$, $h_c(1P)\pi$, $h_c(1P)\eta$,
$\chi_{c0}(1P)\omega(782)$, $\chi_{c2}(1P)\omega(782)$ and
tetraquark states~\cite{Cui:2006mp,Maiani:2014aja,Zhao:2014qva,Wu:2018xdi} and so on. Their masses are
about $\sim(4.0-4.1)~\text{GeV}$. Thus for the higher mass states,
such as $\psi(4660)$, $\psi(4415)$, and $\psi(4360)$, taking
$E_{k}-E_{A}$ as a constant is a reasonable assumption both at quark
and hadron levels. However, for the lower mass states like
$\psi(4040)$ and $\psi(4160)$, the $E_{k}-E_{A}$'s are quite small
and are sensitive to the masses of intermediate state.
%their differences become important at the hadron level.
In this case, taking $E_{k}-E_{A}$ as a constant will introduce
a large uncertainty in our calculation.

\subsection{Bottomonim}
\label{sec:bottomonim}

\begin{table*}[]
\centering \caption{The $B^{(*)}\bar{B}^{(*)}\pi$ partial decay
widths of the vector bottomonium (in units of MeV).
$\mathcal{B}_{\text{exp}}$ represents the branching ratio for each
corresponding channel.} \label{table:width:bb:BBpi}
\begin{tabular}{lclccccccccccccccc}
\toprule[0.5pt]\toprule[0.5pt]
~~~~~~&~~~~~~&~~~~~~&\multicolumn{2}{c}{$\beta=550
\text{MeV}$}&~~~~~~&\multicolumn{2}{c}{$\beta=600
\text{MeV}$}&~~~~~~&\multicolumn{2}{c}{$\beta=650
\text{MeV}$}&~~~~~~&~~~~~~\\ \cline{4-5}\cline{7-8}\cline{10-11}
Meson ~~~~~& State ~~~~~& Mode ~~~~~&
$\Gamma_{\text{th}}$~~~~~&$\mathcal{B}_i$& ~~~~~&
$\Gamma_{\text{th}}$~~~~~&$\mathcal{B}_i$& ~~~~~&
$\Gamma_{\text{th}}$~~~~~&$\mathcal{B}_i$ ~~~~~&
\cite{Simonov:2008cr}
~~~~~& $\mathcal{B}_{\text{exp}}$~\cite{Tanabashi:2018oca} \\
\midrule[0.5pt] $\Upsilon(10860)$ ~~~~~& $5{^3S_1}$ ~~~~~& $BB\pi$
~~~~~& $0.12$ ~~~~~&$0.2\%$&~~~~~& $0.20$ ~~~~~&$0.4\%$&~~~~~&
$0.28$~~~~~&$0.5\%$ ~~~~~&
$$ ~~~~~&$(0.0\pm1.2)\%$ \\
~~~~~&~~~~~& $BB^*\pi$ ~~~~~& $1.36$~~~~~&$2.7\%$& ~~~~~&
$1.22$~~~~~&$2.4\%$& ~~~~~& $0.94$~~~~~&$1.8\%$ ~~~~~&
(23--30)~\text{keV} ~~~~~&
$(7.3\pm2.3)\%$ \\
~~~~~&~~~~~& $B^*B^*\pi$~~~~~ & $0.68$~~~~~&$1.3\%$& ~~~~~&
$0.61$~~~~~&$1.2\%$& ~~~~~& $0.47$~~~~~&$0.9\%$ ~~~~~&
(5--6.6)~\text{keV} ~~~~~&
$(1.0\pm1.4)\%$ \\
\midrule[0.5pt]
$\Upsilon(11020)$ ~~~~~& $6{^3S_1}$ ~~~~~& $BB\pi$ ~~~~~& $0.17$~~~~~&$0.3\%$& ~~~~~& $0.34$~~~~~&$0.7\%$& ~~~~~& $0.55$~~~~~&$1.1\%$ ~~~~~& ~~~~~&\\
~~~~~& ~~~~~& $BB^*\pi$ ~~~~~& $2.50$~~~~~&$5.1\%$& ~~~~~& $3.17$~~~~~&$6.5\%$& ~~~~~& $3.41$~~~~~&$7.0\%$ ~~~~~& ~~~~~& \\
~~~~~& ~~~~~& $B^*B^*\pi$ ~~~~~& $2.12$~~~~~&$4.3\%$& ~~~~~& $2.69$~~~~~&$5.5\%$& ~~~~~& $2.97$~~~~~&$6.1\%$ ~~~~~& ~~~~~&\\
\bottomrule[0.5pt]\bottomrule[0.5pt]
\end{tabular}
\end{table*}

For the bottomonium system, there are three $b\bar{b}$ states above
the open bottom threshold, namely $\Upsilon(4S)$, $\Upsilon(10860)$,
and $\Upsilon(11020)$. A number of studies are available on the
study of their strong decays with the ${^3P_0}$
model~\cite{Segovia:2012cd,Ferretti:2013vua,Godfrey:2015dia} and
other models~\cite{Simonov:2008cr,Ebert:2014jxa}.
 Most of them focus on the two body strong decays.
 However, the $\Upsilon(10860)$ and $\Upsilon(11020)$ states can also decay into two bottomed mesons plus a $\pi$ meson.
Furthermore, these channels for the $\Upsilon(10860)$ state have
recently been observed by the Belle
Collaboration~\cite{Drutskoy:2010an}. We investigate the three
body decays of $\Upsilon(10860)$ and $\Upsilon(11020)$ with the
extended ${^3P_0}$ model.

The $\Upsilon(10860)$ and $\Upsilon(11020)$ were discovered by the
CLEO Collaboration in the $e^+e^-$
annihilation~\cite{Besson:1984bd}. Their masses and widths
are~\cite{Tanabashi:2018oca}
\begin{eqnarray}
m_{\Upsilon(10860)} &=& 10889.9_{-2.6}^{+3.2}~\text{MeV}, \\
\Gamma_{\Upsilon(10860)} &=& 51_{-7}^{+6}~\text{MeV}, \\
m_{\Upsilon(11020)} &=& 10992.9_{-3.1}^{+10.0}~\text{MeV}, \\
\Gamma_{\Upsilon(11020)} &=& 49_{-15}^{+9}~\text{MeV}.
\end{eqnarray}
They are usually assigned to be the $5{^3S_1}$ and $6{^3S_1}$
$b\bar{b}$ states in the quark model. We discuss the three body
decays of the $\Upsilon(10860)$ and $\Upsilon(11020)$ states with
this assignment.

The partial decay widths of the $\Upsilon(10860)$ state are listed
in Table~\ref{table:width:bb:BBpi}. According to our calculation, we
obtain
\begin{eqnarray}
& \Gamma[\Upsilon(10860){\to}BB\pi] \sim 0.20~\text{MeV}, \\
& \Gamma[\Upsilon(10860){\to}BB^*\pi] \sim 1.22~\text{MeV}, \\
& \Gamma[\Upsilon(10860){\to}B^*B^*\pi] \sim 0.61~\text{MeV}.
\end{eqnarray}
The $BB^*\pi$ decay width is the largest one. Combining with the
total width of $\Upsilon(10860)$, we obtain the branching ratios as
follows:
\begin{eqnarray}
& \mathcal{B}[\Upsilon(10860){\to}BB\pi] \sim 0.4\%, \\
& \mathcal{B}[\Upsilon(10860){\to}BB^*\pi] \sim 2.4\%, \\
& \mathcal{B}[\Upsilon(10860){\to}B^*B^*\pi] \sim 1.2\%.
\end{eqnarray}
The predicted branching ratios of the $BB\pi$ and $B^*B^*\pi$ decay
modes are within the ranges of experimental values measured by the
Belle Collaboration~\cite{Tanabashi:2018oca}. For the $BB^*\pi$
decay mode, our result is slightly smaller than the experiment data.

The partial decay widths of $\Upsilon(11020)$ state are also
presented in Table~\ref{table:width:bb:BBpi}. According to our
calculations, we obtain the corresponding branching ratios
\begin{eqnarray}
& \mathcal{B}[\Upsilon(11020){\to}BB\pi] \sim 0.7\%, \\
& \mathcal{B}[\Upsilon(11020){\to}BB^*\pi] \sim 6.5\%, \\
& \mathcal{B}[\Upsilon(11020){\to}B^*B^*\pi] \sim 5.5\%.
\end{eqnarray}
The branching ratios of the $BB^*\pi$ and $B^*B^*\pi$ channels are
quite large. Thus these two channels may be observed by the Belle II
Collaboration in the near future.

\subsection{The effect of $\beta$}
\label{sec:beta}

We have investigated the three body open flavor decays of five
charmoniumlike states with various assignments and two bottomonium
states. In this work, we carried out the calculation by fixing the
harmonic oscillator parameter $\beta$ to be $500~\text{MeV}$
($600~\text{MeV}$) for charmonium (bottomonium) states. However, the
parameter $\beta$ of the initial states is not determined precisely,
which may bring in uncertainty to our results. To estimate this
effect, we carry out the preceding calculation by varying the
parameter $\beta$ of the initial states by $50~\text{MeV}$. We
investigate the decay properties with two different $\beta$ values,
$\beta=450~\text{MeV}$ and $\beta=550~\text{MeV}$ for charmonium
states and $\beta=550~\text{MeV}$ and $\beta=650~\text{MeV}$ for
bottomonium states. The numerical results are presented in
Table~\ref{table:width:cc:4040+4160:DDpi},
Figs.~\ref{fig:DDpi}--\ref{fig:DstarDstarpi} for charmonium states
and Table~\ref{table:width:bb:BBpi} for bottomonium states.

For the charmonium states $\psi(4360)$, $\psi(4415)$, and $\psi(4660)$
and bottomonium states $\Upsilon(10860)$ and $\Upsilon(11020)$, we
notice that within a reasonable range of the parameter $\beta$, our
main predictions and conclusions hold. However, for the $\psi(4040)$
and $\psi(4160)$ states, the decay widths are quite sensitive to
$\beta$. Particularly, for $\psi(4040)$ decaying into $DD^*\pi$,
when $\beta$ changes by $50~\text{MeV}$, the width varies by a
factor of $5$. As pointed out in
Sec.~\ref{sec:charmonium:4040-4160}, the $\psi(4040)$ and
$\psi(4160)$ states are close to the mass of intermediate states,
and taking $E_k-E_A$ as a constant introduces a large
uncertainty in our calculation.

\section{Conclusions}
\label{Sec:Conclusion}

In the present work, we have investigated the OZI-allowed three body
open flavor decays of excited vector charmoniumlike states and
bottomonium states in the framework of the extended ${^3P_0}$ model.
In spite of the large uncertainty caused by the parameter $\gamma$,
it is the first attempt along this direction in literatures to study
this type of decay modes by considering the creation of two light
$q\bar{q}$ pairs from vacuum. Our main results are summarized as
follows.

For the well-established states $\psi(4040)$ and $\psi(4160)$, we
estimate their three body open flavor decay properties with the
assignments $\psi(3{^3S_1})$ and $\psi(2{^3D_1})$, respectively. The
partial decay widths of $\psi(4040)$ should be fairly small (about
several tens keV), and those of $\psi(4160)$ are a little larger,
which are about $0.1~\text{MeV}$ for the $DD\pi$ and $DD^*\pi$
modes.

We also discuss the decay properties of $\psi(4360)$ as a candidate of
$\psi(4{^3S_1})$ or $\psi(3{^3D_1})$. From our calculation, the
partial decay width of $DD^*\pi$ mode can reach up to
$1~\text{MeV}$ in both cases. Thus if $\psi(4360)$ is one of these
states, it may be observed in the $DD^*\pi$ channel.

With the $\psi(4{^3S_1})$ assignment, the $DD^*\pi$ and $D^*D^*\pi$
decay widths of $\psi(4415)$ are larger than $1~\text{MeV}$.
Meanwhile, the $DD\pi$ decay mode is sizable with a width of
$\sim0.38~\text{MeV}$. Our predictions for the branching ratios of
the $DD\pi$ and $DD^*\pi$ channels are within the upper limits
measured by the Belle
Collaboration~\cite{Pakhlova:2007fq,Pakhlova:2009jv}. However,
assigning $\psi(4415)$ to be the $\psi(5{^3S_1})$ or $\psi(3{^3D_1})$
state, we obtain similar decay properties. Thus, to further
determine the inner structure of $\psi(4415)$, more precise
experimental data are needed.

We calculated the three body open flavor decay widths of $\psi(4660)$
with various assignments, $\psi(4S,5S,6S)$ and $\psi(3D,4D,5D)$. In
both cases, its three body decays are dominated by the $DD^*\pi$ and
$D^*D^*\pi$ channels, and the partial decay widths can reach up to
several MeV. Meanwhile, we notice that the $DD\rho$ and $DD\omega$
decay widths of the $D$-wave states are larger than those of the
$S$-wave states. If $\psi(4660)$ turns out to be $\psi(3D)$, its
$DD\rho$ decay width even reaches up to $1.86~\text{MeV}$.

We have also investigated the three body open flavor decays of
$\Upsilon(10860)$ and $\Upsilon(11020)$. The branching ratios of
$\Upsilon(10860)$ decaying into $BB\pi$ and $B^*B^*\pi$ are
consistent with the experimental data, while the $BB^*\pi$ braching
ratio is smaller but very close to the Belle's measurement. For
$\Upsilon(11020)$, the $BB\pi$, $BB^*\pi$, and $B^*B^*\pi$ decay
widths are $0.34$, $3.17$ and
$2.69~\text{MeV}$, respectively. Hopefully the $BB^*\pi$ and
$B^*B^*\pi$ decay modes of the $\Upsilon(11020)$ state will be
observed by the Belle II Collaboration in the very near future.

\section*{Acknowledgments}

X.~Z.~W. and L.~Y.~X. are grateful to L.~Meng and G.~J.~Wang for helpful discussions.
This project is supported by the National Natural Science Foundation
of China under Grants No. 11575008 and 11621131001 and National Key Basic
Research Program of China (Grants No. 2015CB856700). This project is also in
part supported by China Postdoctoral Science Foundation under Grant
No. 2017M620492.

\begin{appendix}
\section{The momentum-space integration}
\label{App:Sec:Integration}

The momentum-space integration
$I_{M_{L_B},M_{L_C},M_{L_D}}^{M_{L_A},mm'}(\mathbf{p})$ reads
\begin{eqnarray}\label{eqn:momentum-space-integration:3}
&& I_{M_{L_B},M_{L_C},M_{L_D}}^{M_{L_A},mm'}(\mathbf{p})\nonumber\\
&=& \int d^3\mathbf{p}_1
\mathcal{Y}_1^m\left(\mathbf{P}_{B}-\mathbf{p}_1\right)
\mathcal{Y}_1^{m'}\left(-\mathbf{P}_{C}-\mathbf{p}_1\right)\nonumber\\
&\times&
\psi_{n_BL_BM_{L_B}}^*\left(\mathbf{p}_1-\kappa_{1}\mathbf{P}_B\right)\nonumber\\
&\times&
\psi_{n_CL_CM_{L_C}}^*\left(-\mathbf{p}_1-\kappa_{2}\mathbf{P}_C\right)\nonumber\\
&\times&
\psi_{n_DL_DM_{L_D}}^*\left(\mathbf{p}_1-\kappa_{32}\mathbf{P}_B+\kappa_{31}\mathbf{P}_C\right)\nonumber\\
&\times& \psi_{n_AL_AM_{L_A}}\left(\mathbf{p}_1\right) ,
\end{eqnarray}
where
\begin{eqnarray}
\kappa_{1}&  = &\frac{m_1}{m_1+m_3}, \\
\kappa_{2}&  = &\frac{m_2}{m_2+m_6}, \\
\kappa_{31}& =& \frac{m_4}{m_4+m_5}, \\
\kappa_{32}& = &\frac{m_5}{m_4+m_5}.
\end{eqnarray}
In our calculation, only the $S$-wave states are considered for the
final states. Thus we can rewrite the momentum-space integration
$I_{M_{L_B},M_{L_C},M_{L_D}}^{M_{L_A},mm'}(\mathbf{p})$ as
$\Pi\left(M_{L_A},m,m'\right)$.

For the decay of the $1S$ state,
\begin{eqnarray}
&&\Pi\left(0,m,m'\right)\nonumber\\
&&= \left(\frac{1}{\pi\beta^2}\right)^{3/4}
\left(\frac{1}{\pi\beta_{B}^2}\right)^{3/4}
\left(\frac{1}{\pi\beta_{C}^2}\right)^{3/4}
\left(\frac{1}{\pi\beta_{D}^2}\right)^{3/4}\nonumber\\
&&\times\exp\left[f\left(\mathbf{P}_B,\mathbf{P}_C\right)\right]\times\left(\frac{\pi}{\lambda_1}\right)^{3/2}\nonumber\\
&&\times\Bigg[\frac{3(-1)^{m}}{8\pi\lambda_1}\delta_{m',-m} +\mathcal{Y}_1^m\left(-\left(\eta-1\right)\mathbf{P}_B+\varpi\mathbf{P}_C\right)\nonumber\\
&&\times\mathcal{Y}_1^{m'}\left(-\eta\mathbf{P}_B+\left(\varpi-1\right)\mathbf{P}_C\right)\Bigg].
\end{eqnarray}
For the decay of the $1D$ state
\begin{eqnarray}
&& \Pi\left(M_{L_A},m,m'\right)\nonumber\\
&&= -\left(\frac{16}{15\sqrt{\pi}}\right)^{1/2}\frac{1}{\beta^{7/2}}
\cdot\left(\frac{1}{\pi\beta_{B}^2}\right)^{3/4}
\left(\frac{1}{\pi\beta_{C}^2}\right)^{3/4}
\left(\frac{1}{\pi\beta_{D}^2}\right)^{3/4}\nonumber\\
&&\times\exp\left[f\left(\mathbf{P}_B,\mathbf{P}_C\right)\right]\left(\frac{\pi}{\lambda_1}\right)^{3/2}\nonumber\\
&&\times\Bigg\{\frac{15}{16\pi\lambda_1^2} \cdot k_{m,m',M_{L_A}}^{(112)}\delta_{m+m',-M_{L_A}}\nonumber\\
&&\qquad+\frac{3}{8\pi\lambda_1} \cdot\bigg[(-1)^{m}\delta_{m',-m}\mathcal{Y}_{2}^{M_{L_A}}\left(\eta\mathbf{P}_B-\varpi\mathbf{P}_C\right)\nonumber\\
&&\qquad\qquad-(-1)^{m}\sqrt{\frac{40\pi}{3}}\mathcal{Y}_1^{m'}\left(-\eta\mathbf{P}_B+\left(\varpi-1\right)\mathbf{P}_C\right)\nonumber\\
&&\qquad\qquad\qquad\times\mathcal{Y}_{1}^{M_{L_A}+m}\left(\eta\mathbf{P}_B-\varpi\mathbf{P}_C\right)\nonumber\\
&&\qquad\qquad\qquad\times\langle1-m;1M_{L_A}+m|2M_{L_A}\rangle\nonumber\\
&&\qquad\qquad -(-1)^{m'}\sqrt{\frac{40\pi}{3}}\mathcal{Y}_1^m\left(-\left(\eta-1\right)\mathbf{P}_B+\varpi\mathbf{P}_C\right)\nonumber\\
&&\qquad\qquad\qquad\times\mathcal{Y}_{1}^{M_{L_A}+m'}\left(\eta\mathbf{P}_B-\varpi\mathbf{P}_C\right)\nonumber\\
&&\qquad\qquad\qquad\times\langle1-m';1M_{L_A}+m'|2M_{L_A}\rangle\bigg]\nonumber\\
&&\qquad+\mathcal{Y}_1^m\left(-\left(\eta-1\right)\mathbf{P}_B+\varpi\mathbf{P}_C\right)\nonumber\\
&&\qquad\qquad\times\mathcal{Y}_1^{m'}\left(-\eta\mathbf{P}_B+\left(\varpi-1\right)\mathbf{P}_C\right)\nonumber\\
&&\qquad\qquad\times\mathcal{Y}_{2}^{M_{L_A}}\left(\eta\mathbf{P}_B-\varpi\mathbf{P}_C\right)\Bigg\}.
\end{eqnarray}
Here,
\begin{eqnarray}
\lambda_1&=&\frac{1}{2\alpha^2}+\frac{1}{2\alpha_{B}^2}+\frac{1}{2\alpha_{C}^2}+\frac{1}{2\alpha_{D}^2}, \\
\lambda_2&=&\frac{\kappa_{1}}{\alpha_{B}^2}+\frac{\kappa_{32}}{\alpha_{D}^2},\\
 \lambda_3&=&\frac{\kappa_{2}}{\alpha_{C}^2}+\frac{\kappa_{31}}{\alpha_{D}^2};
\end{eqnarray}
\begin{eqnarray}
f\left(\mathbf{P}_B,\mathbf{P}_C\right)& =& \frac{\left(\lambda_2\mathbf{P}_B-\lambda_3\mathbf{P}_C\right)^2}{4\lambda_1}-\frac{\kappa_{1}^2\mathbf{P}_B^2}{2\alpha_{B}^2}\nonumber\\
&&- \frac{\kappa_{2}^2\mathbf{P}_C^2}{2\alpha_{C}^2}- \frac{\left(\kappa_{32}\mathbf{P}_B-\kappa_{31}\mathbf{P}_C\right)^2}{2\alpha_{D}^2},\\
 \eta &=& \frac{\lambda_2}{2\lambda_1},\\
 \varpi &=& \frac{\lambda_3}{2\lambda_1}.
\end{eqnarray}
When $m+m'+M_{L_A}=0$, $k_{m,m',M_{L_A}}^{(112)}$'s are non-vanishing and take the following values,
\begin{eqnarray}
&& k_{\pm1,\pm1,\mp2}^{(112)}
= \sqrt{\frac{3}{10\pi}}, \\
&& k_{\pm1,0,\mp1}^{(112)}=k_{0,\pm1,\mp1}^{(112)}
= -\sqrt{\frac{3}{20\pi}}, \\
&& k_{\pm1,\mp1,0}^{(112)}
= \frac{1}{\sqrt{20\pi}}, \\
&& k_{0,0,0}^{(112)}
= \frac{1}{\sqrt{5\pi}}.
\end{eqnarray}

Applying the above momentum space integrations to
Eq.~(\ref{eqn:amplitude}), we can calculate the $1S$ and $1D$
amplitudes. Amplitudes of the radially and orbitally excited states
can be obtained by following recursion
relations~\cite{Liu:2011yp,Xiao:2018iez},
\begin{eqnarray}
\mathcal{M}_{3S} &=&\sqrt{\frac{2}{15}}\left(
\alpha^2\frac{\partial^2}{\partial\alpha^2}
+\alpha\frac{\partial}{\partial\alpha} +\frac{3}{2}\right)
\mathcal{M}_{1S}, \\
\mathcal{M}_{4S} &=& \frac{2}{3\sqrt{35}} \bigg(
\alpha^3\frac{\partial^3}{\partial\alpha^3}
+3\alpha^2\frac{\partial^2}{\partial\alpha^2}
+\frac{15}{2}\alpha\frac{\partial}{\partial\alpha} \bigg)
\mathcal{M}_{1S}, \\
\mathcal{M}_{5S} &=& \frac{2}{9\sqrt{70}} \bigg(
\alpha^4\frac{\partial^4}{\partial\alpha^4}
+6\alpha^3\frac{\partial^3}{\partial\alpha^3}
+24\alpha^2\frac{\partial^2}{\partial\alpha^2}\nonumber\\
&&+18\alpha\frac{\partial}{\partial\alpha}
+\frac{63}{4}\bigg)\mathcal{M}_{1S}, \\\mathcal{M}_{6S}
&=&\frac{2}{45\sqrt{77}}\bigg(\alpha^5\frac{\partial^5}{\partial\alpha^5}
+10\alpha^4\frac{\partial^4}{\partial\alpha^4}
+60\alpha^3\frac{\partial^3}{\partial\alpha^3}\nonumber\\
&&+120\alpha^2\frac{\partial^2}{\partial\alpha^2}
+\frac{675}{4}\alpha\frac{\partial}{\partial\alpha}\bigg)\mathcal{M}_{1S};
\end{eqnarray}
\begin{eqnarray}
\mathcal{M}_{2D} &=&\sqrt{\frac{2}{7}}
\alpha\frac{\partial}{\partial\alpha}
\mathcal{M}_{1D}, \\
\mathcal{M}_{3D} &=&\frac{1}{3}\sqrt{\frac{2}{7}}
\left(\alpha^2\frac{\partial^2}{\partial\alpha^2}+\alpha\frac{\partial}{\partial\alpha}+\frac{7}{2}\right)
\mathcal{M}_{1D}, \\
\mathcal{M}_{4D} &=&\frac{2}{3\sqrt{231}}
\left(\alpha^3\frac{\partial^3}{\partial\alpha^3}+3\alpha^2\frac{\partial^2}{\partial\alpha^2}+\frac{27}{2}\alpha\frac{\partial}{\partial\alpha}\right)
\nonumber\\
&&\times\mathcal{M}_{1D}, \\
\mathcal{M}_{5D} &=&\frac{1}{3}\sqrt{\frac{2}{3003}}
\bigg(\alpha^4\frac{\partial^4}{\partial\alpha^4}
+6\alpha^3\frac{\partial^3}{\partial\alpha^3}
+36\alpha^2\frac{\partial^2}{\partial\alpha^2}\nonumber\\
&&+30\alpha\frac{\partial}{\partial\alpha}+\frac{231}{4}\bigg)\mathcal{M}_{1D}.
\end{eqnarray}

\end{appendix}

%\end{spacing}

\end{document}